\documentclass[twocolumn]{webofc}
\usepackage[varg]{txfonts}   % Web of Conferences font
\begin{document}
\title{Toroidal dipole mode in nuclei}
%effect of the large axial quadrupole deformation}

\author{
\firstname{Valentin} \lastname{Nesterenko}\inst{1,2}\fnsep\thanks{\email{nester@theor.jinr.ru}}
       \and
        \firstname{Petr} \lastname{Vishnevskiy}\inst{1}\fnsep
          \and
\firstname{Anton} \lastname{Repko}\inst{3}\fnsep
             \and
  \firstname{Paul-Gerhard} \lastname{Reinhard}\inst{4}\fnsep
   \and
  \firstname{Jan} \lastname{Kvasil}\inst{5}\fnsep}

\institute{Laboratory of Theoretical Physics,
  Joint Institute for Nuclear Research, Dubna, Moscow region, 141980, Russia
\and
   Federal State University "Dubna", Dubna, Moscow region, 141980, Russia
\and
 Institute of Physics, Slovak Academy of Sciences, 84511, Bratislava, Slovakia
\and
 Institut f\"ur Theoretische Physik II, Universit\"at Erlangen, D-91058, Erlangen, Germany
 \and
Institute of Particle and Nuclear Physics, Charles University,
CZ-18000, Praha 8, Czech Republic
          }

\abstract{A short review on the toroidal dipole mode (TDM) in nuclei is done.
The appearance of TDM in nuclei is justified. The experimental manifestation of TDM
in $(e,e')$ reaction in $^{58}$Ni is shortly reported. The relation of TDM and pygmy
E1 resonance is discussed.}

\maketitle
\section{Introduction}
\label{intro}
   Dipole toroidal mode (TDM) appears in many physical systems~\cite{dub90}. It is
   well known in hydrodynamics of conventional gases and fluids as so-called Hill's
   vortex, predicted yet in 1894 as a stationary solution of the Euler equations for
   an incompressible fluid~\cite{hil84}. Actually this is a simple case of a turbulent flow
   when the particles exhibit a poloidal current circulating on the
   surface of a torus as illustrated in Fig. 1(a). Smoke rings are a familiar example.
   Various manifestations of TDM are widely discussed (and already observed experimentally)
   in solid-state physics~\cite{dub90},  metamaterials~\cite{kae10} and metaphotonics~\cite{ara20}.
   Besides, TDM was predicted in heavy-ion collisions~\cite{iva23} and  anapole
   dark matter~\cite{anapole}.

 \section{Basics of TDM}

   Toroidal multipoles are fundamental electromagnetic excitations different
from conventional charge and magnetic multipoles~\cite{dub74}. The complete mutipole expansion
of the current density can be done using three independent families of multipoles: irrotational electric,
magnetic and vortical toroidal. This is illustrated by the Chandrasekhar-Moffat
decomposition\cite{dub90,nanz16} of the current density:
\begin{equation}
 \label{eq:dj}
 \delta\bold{j}(\bold{r}) = \bold{\nabla} \phi(\bold{r}) + \bold{\nabla}\times (\bold{r}\psi(\bold{r})) +
 \bold{\nabla}\times \bold{\nabla}\times(\bold{r} \chi(\bold{r})) ,
\end{equation}
%can be done in terms of three independent quantities: irrotational electric, magnetic and vortical toroidal,
%respectively.
where
%($\bold{\nabla}\times \bold{\nabla}\times \bold{\xi}$)
$\phi, \psi, \chi$ are some scalar coordinate-dependent functions often called as Debye potentials.
As compared with familiar Helmholtz's expansion, this decomposition allows to
separate the magnetic $\bold{\nabla}\times (\bold{r}\psi(\bold{r}))$ and electric
$\bold{\nabla}\times \bold{\nabla}\times(\bold{r} \chi(\bold{r}))$ vortical terms.
The last double-curl term produces (after extraction of its long-wave fraction by applying
Siegert theorem) the toridal current \cite{dub90,dub74,kva11}.
In the simplest dipole case, the electric dipole is produced by a pair of opposite charges,
the magnetic dipole is formed by a current loop and the toroidal dipole is generated by currents flowing
on the surface of a torus. The transversal TDM exhibits various exotic electromagnetic properties
which are a subject of extensive studies, see e.g.~\cite{dub90,kae10,ara20,nanz16,afadub98}.
\begin{figure}
\centering
%\sidecaption
\includegraphics[width=8cm]{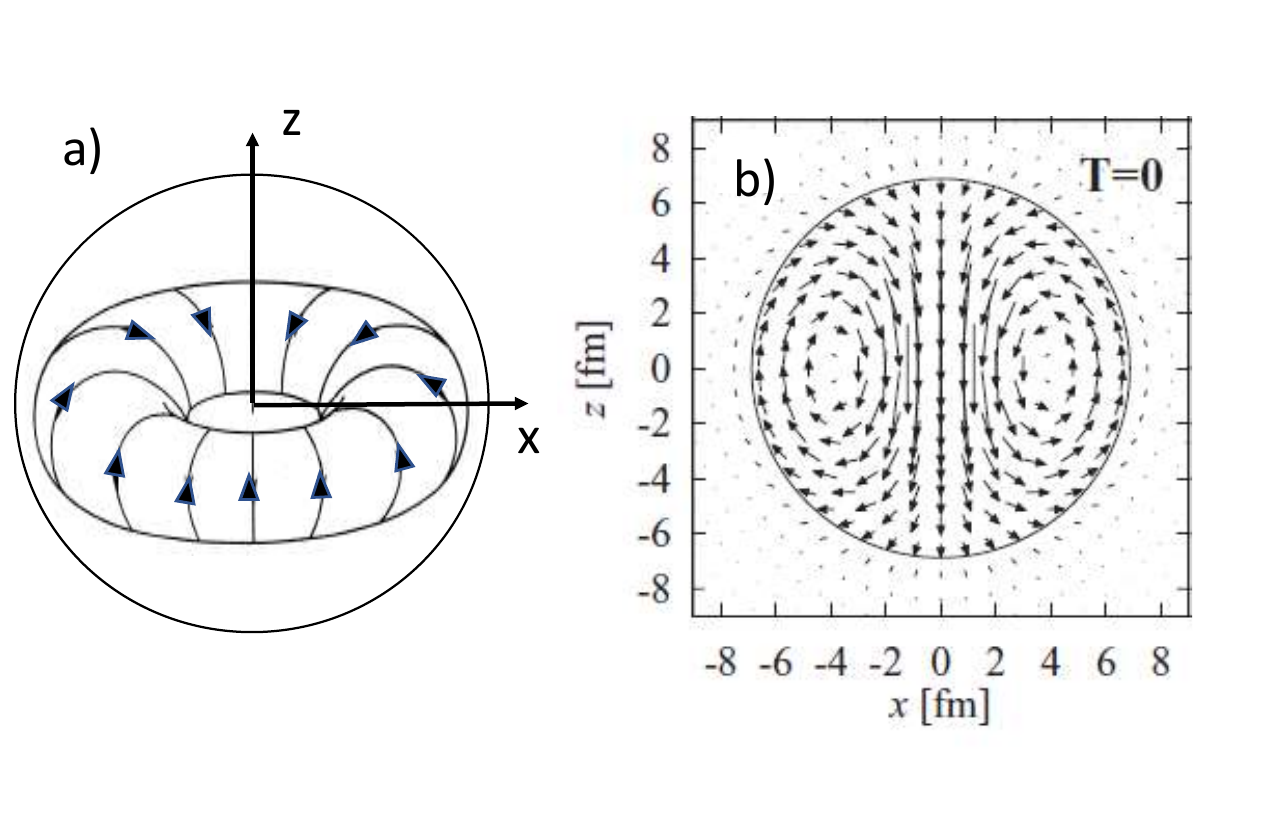}
\caption{(a) Schematic picture of TDM. (b) Isoscalar (T=0) TDM in $^{208}$Pb\cite{rep13}.}
\label{fig-1}       % Give a unique label.
\end{figure}

Since TDM is a general property of many systems, it would be strange if it is absent
in atomic  nuclei. TDM was predicted in nuclei in the seminal paper of Semenko~\cite{sem81}.
Then, it was extensively studied in various theoretical models from macroscopic fluid-dynamical
approaches to microscopic density-functional theory, see the extensive citation in the recent
study~\cite{cos24}. It was justified that TDM should exist in all the nuclei, independently of
their size, Z/N ratio and shape~\cite{kva11,rep19,NePAN16,Ne_PRL18}. Since  TDM is transversal~\cite{kva11},
it does not contribute to nuclear continuity equation (CE). Thus TDM opens a way to explore
yet "terra incognita" of electric nuclear excitations beyond CE.

In Fig. 1(b), the average isoscalar (T=0) transition density of the convective nuclear current for $1^-$
states at 6.0-8.8 MeV in $^{208}$Pb~\cite{rep13} is shown. We see that
the actual nuclear toroidal motion fills in all the nucleus volume with a strong central
flow along the z-axis (defined in an experiment by a beam direction). The stream lines resemble
irregular ellipses. Unlike a classical Hill’s vortex, the nucleons exhibit not full circulations but
small oscillations along the stream lines on the toroidal surface. In general, TDM appears
as a resonance (toroidal dipole resonance  - TDR) embracing many dipole states and
located just at the energy of the pygmy dipole resonance (PDR)~\cite{kva11,rep19,NePAN16}. Besides,
there can exist low-energy individual toroidal states (ITS) lying below PDR~\cite{cos24,Ne_PRL18}.

Our calculations show that the vortical toroidal flow has a mean-field origin. This flow is
mainly produced by particular particle-hole (1ph) $\Delta \cal N$=1 ($\cal N$ is a principle shell quantum number)
dipole excitations. Initially, the bump of 1ph $1^-$  excitations has a major irrotational and minor
vortical (toroidal) fractions. The irrotational  isovector (IV) residual interaction
upshifts most of the irrotational E1 strength and forms the giant dipole resonance (GDR).
%However, this interaction does not attach so much the vortical fraction. So,
After removal of the dominant irrotatation fraction from the initial 1ph bump, this bump becomes
mainly toroidal. Following this scheme, TDM is a general feature of atomic nuclei.

\begin{figure} % fig. 2
\centering
%\sidecaption
\includegraphics[width=8cm]{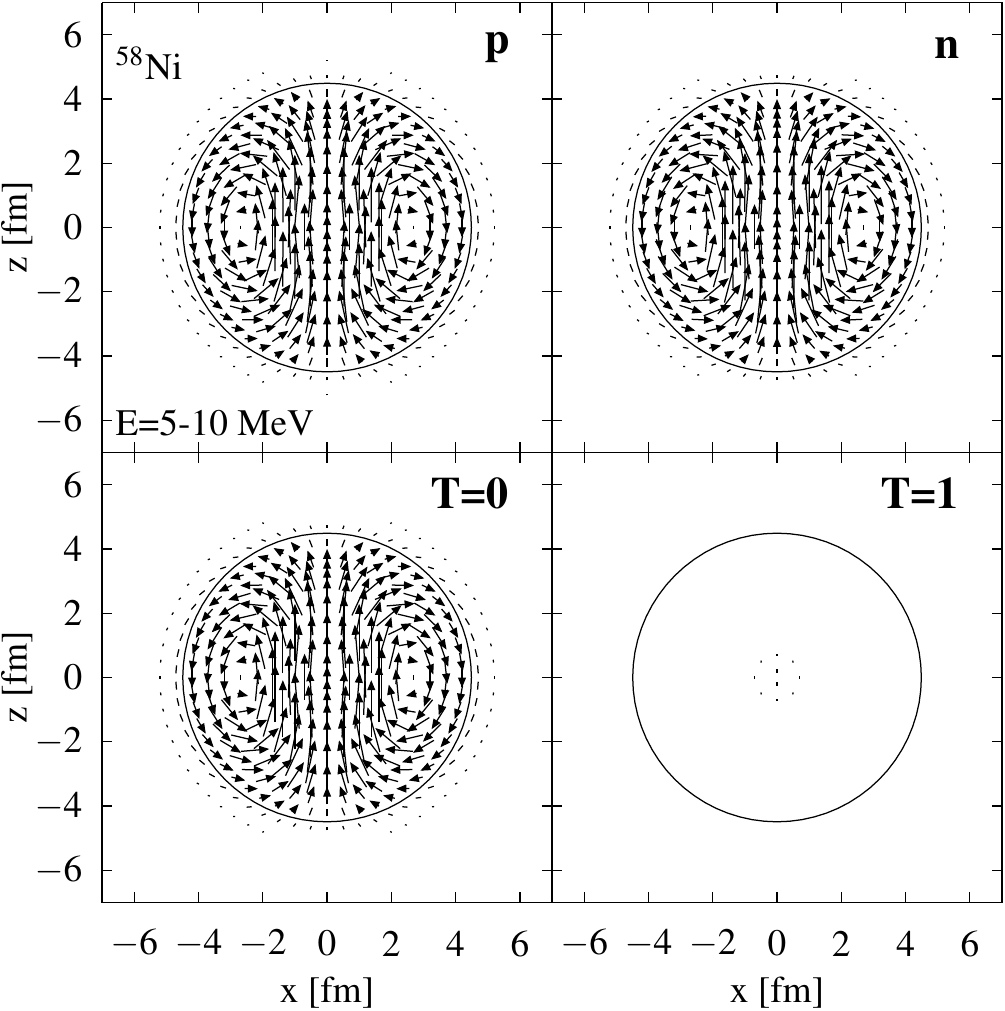}
\caption{Proton (p), neutron (n), IS (T=0) and IV (T=1) convective current transition densities
averaged for QRPA electric dipole states at 5-10 MeV in $^{58}$Ni. The calculations are
performed with Skyrme parametrization  SVmas10.}
%Please write your figure caption here}
\label{fig-2}       % Give a unique label
\end{figure}

The IS dipole residual interaction downshifts the toroidal strength and makes it
stronger~\cite{kva11,rep19,NePAN16}.   The resulting IS TDM is located below GDR
and so is most suitable for the experimental observation. IS TDM has been already observed in numerous
experiments with IS reaction $(\alpha, \alpha')$, see e.g. refs.~\cite{uchi04,bra19}.
However it was a challenge to observe and identify TDM in other reactions, for example,
in $(e,e')$~\cite{Ne_PRC19}.

\section{TDM in $^{58}$Ni}

In early $(e,e')$  experiments performed in TU Darmstadt, at least six dipole states  were found
at 5-10 MeV in spherical nucleus $^{58}$Ni~\cite{met87,Reitz}. These states demonstrated the
enhanced slope of transversal form factors (TFF) at back scattering angles and so were
initially assigned as M1 excitations. However, the subsequent $({\vec\gamma},\gamma')$
experiments have shown that these states are electric dipoles~\cite{bau00,sch13}. Then
there arose a question on the origin of the high TFF. The recent comprehensive study~\cite{cos24}
has shown that dipole states at 5-10 MeV in $^{58}$Ni are IS toroidal and just their toroidal nature
explains the enhanced slope of TFF in $(e,e')$ experiment~\cite{met87}. Instead, the irrotational E1 states
(like GDR and compression mode) do not provide the necessary slope.

The calculations~\cite{cos24} were performed within fully self-consistent Quasiparticle Random-Phase
Approximation (QRPA)~\cite{Ben03,Repcode} with various Skyrme forces. The $(e,e')$ cross-section was computed
in Plane-Wave Born Approximation (PWBA). The toroidal nature of the states at 5-10 MeV was justified
by their high response to the transition toroidal operator and toroidal-like distributions of the
current transition densities (CTD)~\cite{cos24}. Thus, for the first time, the individual
low-energy toroidal dipole states were identified in $(e,e')$ reaction.

 \begin{figure} % fig. 3
\centering
%\sidecaption
\includegraphics[width=9cm]{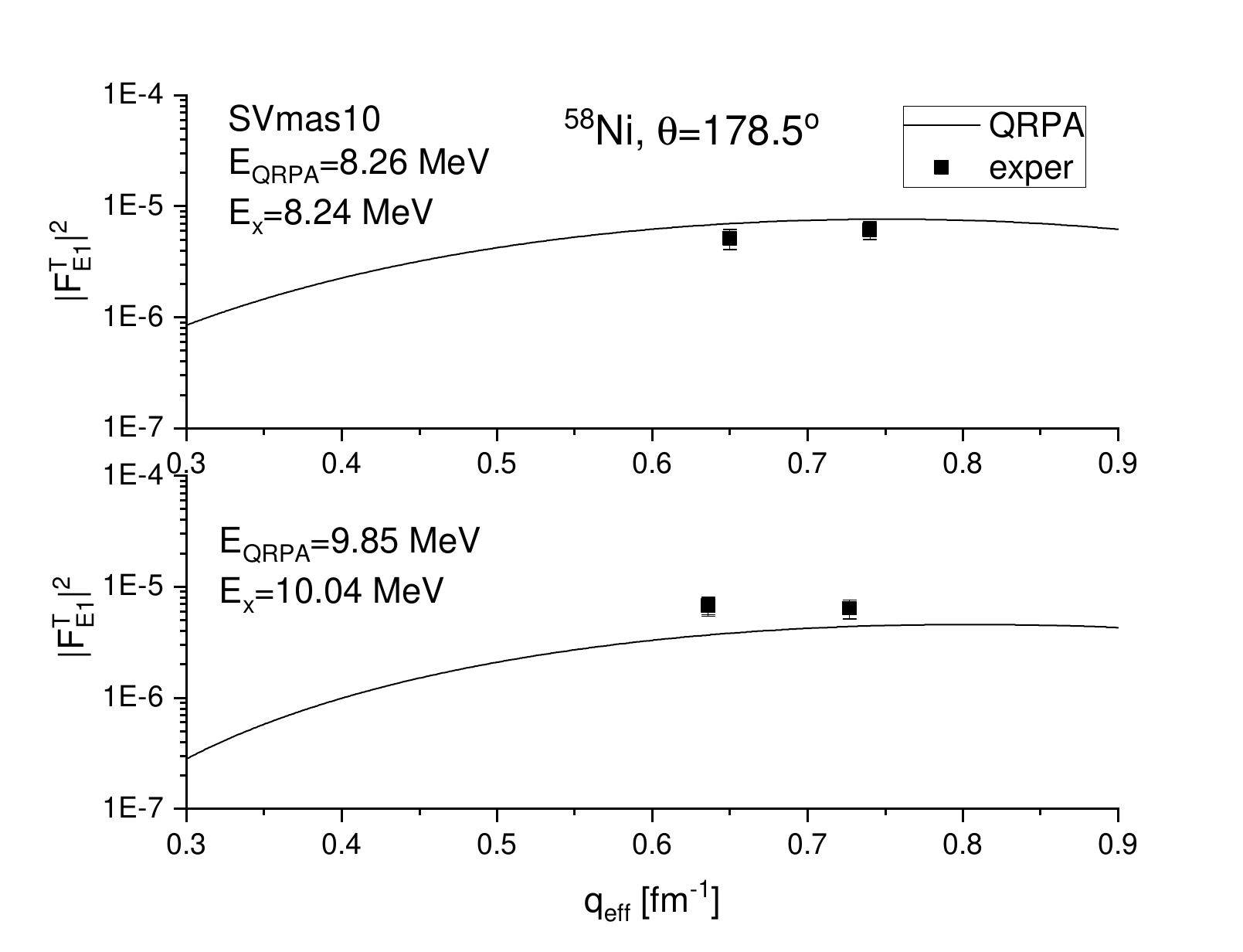}
\caption{The squared TFF computed with Skyrme parametrization SVmas10
for QRPA states at 8.26 and 9.85 MeV, as compared
with experimental data~\cite{Reitz} for the states at 8.24 and 10.04 MeV.}
%Please write your figure caption here}
\label{fig-3}       % Give a unique label
\end{figure}

In Fig.~\ref{fig-2}, the average convective CTD for QRPA E1 states  at 5-10 MeV in $^{58}$Ni
are demonstrated. The calculations were performed with Skyrme force SVmas10~\cite{Klu09}.
Due to the large IS effective mass $m^*/m$=1.0, this force provides a reasonable single-particle
spectra and essential IS residual interaction, which altogether allows to reproduce well the energies
of individual E1 states at 5-10 MeV in $^{58}$Ni~\cite{cos24}. The subsequent CTD averaging
allows to suppress CTD peculiarities of individual states
and exhibit their common average features. The averaging technique is described in Ref.~\cite{rep13}.

The comparison of Figs.~\ref{fig-1} and \ref{fig-2} shows that both proton  and neutron average
currents are obviously toroidal. The total flow is IS, which denotes the IS character of
the considered electric dipole states. The compression E1 mode at 5-10 MeV was
shown to be negligible~\cite{cos24}.

In the study~\cite{cos24}, the  experimental $(e,e')$ data of W. Mettner~\cite{met87}
for scattering angles $\theta =92.9^{\rm o}-164.9^{\rm o}$ were used. It would be also
interesting to compare QRPA results using SVmas10 with additional $(e,e')$
experimental data of B. Reitz~\cite{Reitz}, obtained  at very large scattering angle $\theta =178.5^{\rm o}$.
At so large angle, the contribution of the Coulomb form factor is negligible and so the direct comparison
of the calculated and experimental TFF is possible. In Fig.~\ref{fig-3}, we compare TFF for QRPA states at
8.26 and 9.85 MeV and neigbouring observed states at 8.24 and 10.04 MeV. We see a reasonable agreement
between the theory and experiment, which, additionally to exploration~\cite{cos24}, confirms the validity
of our QRPA  wave functions.
Then the toroidal flow obtained with these wave functions can also be considered as a reliable result.

\section{Discussion}

As mentioned above, SVmas10 gives a reasonable energy spectrum of low-lying E1 states
in $^{58}$Ni~\cite{cos24}. However, this force essentially underestimates reduced transition
probabilities $B(E1)$ of some states. In particular, the calculations give
$B(E1)=6.2 \cdot 10^{-5} {\rm e^2 fm^2}$ for QRPA 8.26-MeV state, which is much smaller
of the experimental value  $18.5 \cdot 10^{-3} {\rm e^2 fm^2}$ obtained for the neighbouring
observed 8.24-MeV state.  It seems that the strong IS residual interaction from SVmas10
makes low-lying E1 states too isoscalar (see Fig.~\ref{fig-2}, where
the nuclear current is fully IS) and so drastically decreases $B(E1)$ values. Perhaps, we need here
the coupling with complex configurations (CCC) which, in principle, can downshift E1
states but, at the same time, keep a sufficiently large IV fraction of the states,
thus providing  more reasonable  $B(E1)$ values~\cite{bau00,sch13}.

Note that the irrotational  fraction of E1 states determines $B(E1)$-value but almost does not affect
TFF at large scattering angles $\theta$~\cite{cos24}.  Instead, the vortical toroidal
fraction of E1 states does not contribute to $B(E1)$  but is crucial in description of
TFF at large $\theta$. CCC can redistribute  the irrotational and toroidal fractions
between concrete states but cannot demolish the average dominance of
the toroidal flow at 5-10 MeV (demonstrated in Fig.~\ref{fig-2}) and a
good description of $(e,e')$ data~\cite{met87,Reitz}. So our main conclusion
on the dominant toroidal dipole mode at 5-10 MeV in $^{58}$Ni should remain valid even after
inclusion of CCC. Moreover, our results show that assignment and treatment of numerous early $(e,e')$
experimental data for low-energy magnetic states should be revisited. Perhaps, some of these
states are actually E1 toroidal.

The nucleus $^{58}$Ni has a minimal neutron excess and so PDR here is absent. Besides, the
analysis~\cite{cos24} was done mainly for E1 states below a typical PDR energy region
$E \approx$ (50 - 60) $A^{-1/3}$ MeV. However, numerous studies performed within different models
show that E1 states at the PDR energy region are also basically toroidal, see
e.g.~\cite{rep13,kva11,rep19,rye02,vre02}. In nuclei with a neutron excess, TDR demonstrates
the same nucleon transition densities as PDR. However, the calculated CTD at the PDR energy region
clearly show the vortical toroidal flow instead of the irrotational PDR-like motion~\cite{rep13,rep19}.
So a familiar PDR picture (oscillation of the neutron excess against Z=N core)
is a rough simplification of the actual nuclear flow.

However, PDR $B(E1)$ values are usually
calculated without application of the simplified PDR-scheme. {\it So,
%the acceptance of the  dominant toroidal flow  does not mean that
there is no any need in revision of the previous numerical B(E1) PDR results
and their implementation for astrophysical applications and construction of the equation of state.}
E1 states at PDR energy region are complicated and have two (irrotational and vortical)
fractions which manifest themselves by a different manner. The dominant toroidal flow is decisive
for description of TFF and $(e,e')$ cross section at large scattering angles.
However, we may forget about this flow if we are interested in calculation
of E1 strength which is determined by the minor
irrotation fraction (including GDR tail etc) of the states.

Finally, it is worth to list main points of interest for TDM.
1) By exploring TDM, we get the actual flow of nucleons in low-energy E1 states instead of
oversimplified  PDR picture.
2) TDM is a new kind of nuclear dynamics beyond CE. The
experimental search of TDM will allow to develop new experimental techniques for excitation and
discrimination of electric vortical states in  nuclei.
3) TDM is a fundamental mode as it represents a third family of electromagnetic form factors,
in addition to familiar magnetic and irrotational electric ones~\cite{dub74,nanz16}. TDM is
related to Zeldovich anapole~\cite{zel}.
4) As mentioned above, TDM is known in solid state physics and other systems~\cite{dub90}. This mode has some
unusual electromagnetic radiative properties~\cite{afadub98} and so is now actively explored in
metaphotonics~\cite{ara20}  and metamaterials~\cite{kae10}. TDM in atomic nuclei provide a unique
chance to explore independently specific properties of the toroidal excitations and compare
them with those  in other physical systems.

\acknowledgement
{We thank Prof. P. von Neumann-Cosel for providing the experimental $(e,e')$ data.
A.R. acknowledges support by the  Slovak Research and Development Agency under Contract
No. APVV-24-0516  and by the Slovak grant agency VEGA (Contract No. 2/0175/24.
J. K. appreciates the support by a grant of the Czech Science Agency No. 23-06439S.}
%19-14048S.}

\end{document}